\documentclass{article}
\usepackage{spconf}
\ninept
\newcommand{\spaceup}{\vspace{-2.3mm}}
\usepackage{lipsum}

\usepackage[utf8]{inputenc}
\usepackage[cmex10]{amsmath}
\usepackage{amssymb}
\usepackage{amsthm}
\usepackage{wasysym}
\usepackage{balance}
\usepackage{url}

\usepackage{graphicx}
\usepackage{epstopdf}
\usepackage{tikz}
\usepackage{tikz-3dplot}
\usetikzlibrary{fit}
\usepackage{pgfplots}
 \usepgfplotslibrary{groupplots}
 \usepgfplotslibrary{external}
\usepackage{textcomp}

\usepackage{color}
\usepackage{subcaption}
\usepackage{bm}

\def\kth{$k^{\text{th}}$}
\def\ith{$i^{\text{th}}$}	
\def\jth{$j^{\text{th}}$}
\renewcommand{\vec}[1]{\mathbf{#1}}

\newcommand{\be}{\begin{equation}}

\newcommand{\bc}{\begin{center}}
\newcommand{\bfl}{\begin{flushleft}}

\newcommand{\beqa}{\begin{eqnarray}}
\newcommand{\beqan}{\begin{eqnarray*}}
\newcommand{\beq}{\begin{equation}}

\renewcommand*\env@matrix[1][*\c@MaxMatrixCols c]{%
  \hskip -\arraycolsep
  \let\@ifnextchar\new@ifnextchar
  \array{#1}}
  
\usepackage{etoolbox}
\preto{\section}{\setcounter{subsubsection}{0}}
\usepackage{booktabs}       
\title{Sliding Bidirectional Recurrent Neural Networks \\for Sequence Detection in Communication Systems}
\name{Nariman Farsad and Andrea Goldsmith\thanks{This work was funded by NSF Center for Science of Information
		grant NSF-CCF-0939370.}}

\address{Electrical Engineering, Stanford University, Stanford, CA}

\begin{document}
\topmargin=0mm
%
\maketitle
\begin{abstract}
The design and analysis of communication systems typically rely on the development of mathematical models that describe the underlying communication channel. However, in some systems, such as molecular communication systems where chemical signals are used for transfer of information, the underlying channel models are unknown. In these scenarios, a completely new approach to design and analysis is required. In this work, we focus on one important aspect of communication systems, the detection algorithms, and demonstrate that by using tools from deep learning, it is possible to train detectors that perform well without any knowledge of the underlying channel models. We propose a technique we call sliding bidirectional recurrent neural network (SBRNN) for real-time sequence detection. We evaluate this algorithm using experimental data that is collected by a chemical communication platform, where the channel model is unknown and difficult to model analytically. We show that deep learning algorithms perform significantly better than a detector proposed in previous works, and the SBRNN outperforms other techniques considered in this work.
\end{abstract}


\section{Introduction}
\label{sec:intro}
The design and analysis of communication systems has relied on developing mathematical models that describe signal transmission, signal propagation, receiver noise, and many other components of the system that affect the end-to-end signal transmission. Particularly, most communication systems today lend themselves to tractable channel models based on a simplification of Maxwell's electromagnetic (EM) equations. However, there are cases where this does not hold, either because the EM signal propagation is very complicated and/or poorly understood, or because the signal is not an EM signal and good models for its propagation don't exist. 
Some examples of the latter includes underwater communication using acoustic signals \cite{sto09}, and a new technique called molecular communication, which relies on chemical signals to interconnect tiny devices with sub-millimeter dimensions in environments such as inside the human body \cite{mor06,aky08,eckBook,far16ST}. In these scenarios, a new approach to design and engineer these systems that does not require analytical channel models is required. 



Motivated by the recent success of deep learning in speech and image processing, where modeling can be difficult, we consider using deep learning in design and analysis of communication systems \cite{lec15,goodfellowBook,ibn00}. Some examples of machine learning tools applied to design problems in communication systems include multiuser detection in code-division multiple-access (CDMA) systems \cite{aaz92,mit94,jua06,isi07}, decoding of linear codes \cite{nac16}, design of new modulation and demodulation schemes \cite{osh16}, and estimating channel model parameters \cite{lee17}. Most previous works have used machine learning to improve one component of the communication system {\em based on the knowledge of the underlying channel models}. 

Our approach is different from these works since we assume that the mathematical models for the communication channel are {\em completely unknown}. Particularly, we focus on one of the important modules of any communication system, the detection algorithm, where the transmitted signal is estimated from a noisy and corrupted version that is observed at the receiver. We demonstrate that, using known deep learning architectures such as a recurrent neural network (RNN), it is possible to train a detector without any knowledge of the underlying system models. Particularly, we use an experimental platform for molecular communication presented in \cite{far17Expt} for generating data for training and testing. 
We also propose a real-time deep learning detector, which we call the {\em sliding bidirectional RNN (SBRNN) detector}, that detects the symbols in an incoming data stream using a dynamic programming approach. This technique could be extended to any type of real-time estimation of sequences in data streams. We demonstrate our SBRNN performs better than other deep learning detectors considered in this work, and significantly better than a detector used in  \cite{far13,koo16}.


\section{Problem Statement}
\label{sec:probState}
In a digital communication system data is converted into a sequence of transmission symbols. Let $\mathcal{X}=\{s_1, s_2, \cdots, s_m\}$ be a finite set of all transmission symbols, and $x_k\in\mathcal{X}$ be the \kth~transmission symbol. The transmission symbols are converted into transmission signals using different modulation techniques, and the signal then propagates in the environment until it arrives at the receiver. The signal that is observed at the destination is a noisy and corrupted version due to the perturbations that are introduced as part of transmission, propagation, and reception processes. Let the random vector $\vec{y}_k$ of length $\ell$ be the observed signal at the destination during the \kth~transmission. Note that the observed signal $\vec{y}_k$ is typically a vector while the transmission symbol is typically a scalar. A {\em detection algorithm} is then used to estimate the transmission symbols from the observed signal at the receiver. Let $\hat{x}_k$ be the symbol estimate for the \kth~transmission. After detection, the estimated transmission symbols are passed to a channel decoder to correct some of the errors in detection. 

Typically, to design the detection algorithm, mathematical channel models are required. These models describe the relation between the transmitted symbols and the received signal through the probabilistic model:
\begin{align}
	P(\vec{y}_1, \vec{y}_{2}, \cdots \mid x_1, x_2, \cdots;\mathbf{\Theta}),
\end{align} 
where $\mathbf{\Theta}$ are the model parameters or the channel state information (CSI). The detection can be performed through symbol-by-symbol detection where $\hat{x}_k$ is estimated from $\vec{y}_k$, or using sequence detection where the sequence $\hat{x}_k, \hat{x}_{k-1}, \cdots, \hat{x}_1$ is estimated from the sequence $\vec{y}_k, \vec{y}_{k-1}, \cdots, \vec{y}_1$. 

An important open problem is the best approach to designing detection algorithms when the underlying channel models are complex such that they cannot provide any insight, or are partly or completely unknown. There are also scenarios where even when the channel models are known, the optimal detection algorithm or even heuristic algorithms can be computationally complex. For example, the complexity of the Viterbi algorithm used in communication channels with memory increases exponentially with memory length, and quickly becomes infeasible for systems with long memory.

Motivated by the recent success of deep learning in speech and image processing, where the underlying models are complex \cite{hin12,lec15}, we propose a data driven approach for decoding based on deep learning.  

\section{Detection Using Deep Learning}
\label{sec:deepDetect}
Estimating the transmitted symbol from the received signals $\vec{y}_k$ can be performed through supervised learning. Particularly, let $m=|\mathcal{X}|$ be the total number of symbols, and let $\vec{p}_k$ be the one-hot representation of the symbol transmitted during the \kth~transmission. Therefore, the element corresponding to the symbol that is transmitted is 1, and all other elements of $\vec{p}_k$ are zero. Note that this is also the PMF of the transmitted symbol during the \kth~transmission where at the transmitter, with probability 1, one of the symbols is transmitted. Also note that the length of the vector $\vec{p}_k$ is different from the length of the vector of the observation signal $\vec{y}_k$ at the destination.

The detection algorithm goes through two phases. In the first phase, known sequences of transmission symbols are transmitted repeatedly and received by the system to create a set of training data. The data can be generated using mathematical models, simulations, experimental measurements, or field measurements. Let $\vec{P}_K = [\vec{p}_1,\vec{p}_2,\cdots,\vec{p}_K]$ be a sequence of $K$ consecutively transmitted symbols, and $\vec{Y}_K = [\vec{y}_1,\vec{y}_2,\cdots,\vec{y}_K]$ the corresponding sequence of observed signals at the destination. Then, the training dataset is represented by
\begin{align}
	\{(\vec{P}^{(1)}_{K_1},\vec{Y}^{(1)}_{K_1}),(\vec{P}^{(2)}_{K_2},\vec{Y}^{(2)}_{K_2}), \cdots, (\vec{P}^{(n)}_{K_n},\vec{Y}^{(n)}_{K_n}) \},
\end{align}
which consists of $n$ samples, and \ith~sample has $K_i$ consecutive transmissions. 

This dataset is then used to train a deep learning classifier that classifies the received signal $\vec{y}_k$ as one of the transmission symbols in $\mathcal{X}$. The input to the deep learning network can be the raw observed signals $\vec{y}_k$, or a set of features $\vec{r}_k$ extracted from the received signals. The output of the deep learning network are the vectors $\hat{\vec{p}}_k$, which are the estimated PMFs that the \kth~transmission symbol belongs to each of the $m$ possible symbols. Note that this output is also useful for soft decision channel decoders (i.e., decoders where the decoder input are PMFs), which are typically the next module after detection in communication systems. If channel coding is not used, the symbol with the highest mass point in $\hat{\vec{p}}_k$ is chosen as the estimated symbol for the \kth~transmission. 

During the training, an optimization algorithm such as stochastic gradient descent is used to minimize the loss between the actual PMF $\vec{p}_k$, and the estimated PMF $\hat{\vec{p}}_k$ \cite{goodfellowBook}. Particularly, the cross-entropy loss function can be used for this optimization \cite{goodfellowBook}. Note that minimizing this loss function is equivalent to minimizing the Kullback-Leibler divergence between the true PMF and the one estimated based on the neural network. It is also equivalent to maximizing the log-likelihood of the correctly  transmitted symbol. 
Therefore, deep learning can be a powerful tool for designing detection algorithms for communication systems, especially when the underlying channel models are unknown. We now discuss how several well-known NN architectures can be used for sequence detection.

\subsection{Sequence Detection}
Sequence detection can be performed using recurrent neural networks (RNN) \cite{lec15,goodfellowBook}, which are well established for sequence estimation in different problems such as  neural machine translation \cite{bah14}, speech recognition \cite{hin12}, or bioinformatics \cite{li16}. In particular,  in this work we use long short-term memory (LSTM) networks \cite{hoc97}. 
One of the main benefits of this detector is that after training, similar to a symbol-by-symbol detector, it can perform detection on any data stream as it arrives at the receiver.  Note that in this configuration the observed signal during the \jth~transmission slot, $\vec{y}_j$ where $j>k$, may carry information about the \kth~symbol $x_k$ due to the ISI. However, since RNNs are feed-forward only, during the estimation of $\hat{x}_k$, the observation signal $\vec{y}_j$ is not considered.  

One way to overcome this limitation is by using bidirectional RNNs (BRNNs) \cite{sch97}. Particularly, we use a bidirectional LSTM (BLSTM) network \cite{gra05} in this work, where the sequence of received signals are once fed in the forward direction into one LSTM cell, and once fed in backwards into another LSTM cell. The two outputs may be passed to more bidirectional layers. This BLSTM architecture ensures that in the estimation of a symbol, future signal observations are taken into account. The main limitation is that as signals from a data stream arrive at the destination, the block length increases and the whole block needs to be detected altogether again for each new data symbol that arrives at the destination. Therefore, this quickly becomes infeasible for long data streams. In the next section we present a new technique to solve this issue.

\subsection{Sliding BRNN Detector}

\begin{figure}
	\centering
	\includegraphics[width=0.8\columnwidth,keepaspectratio]{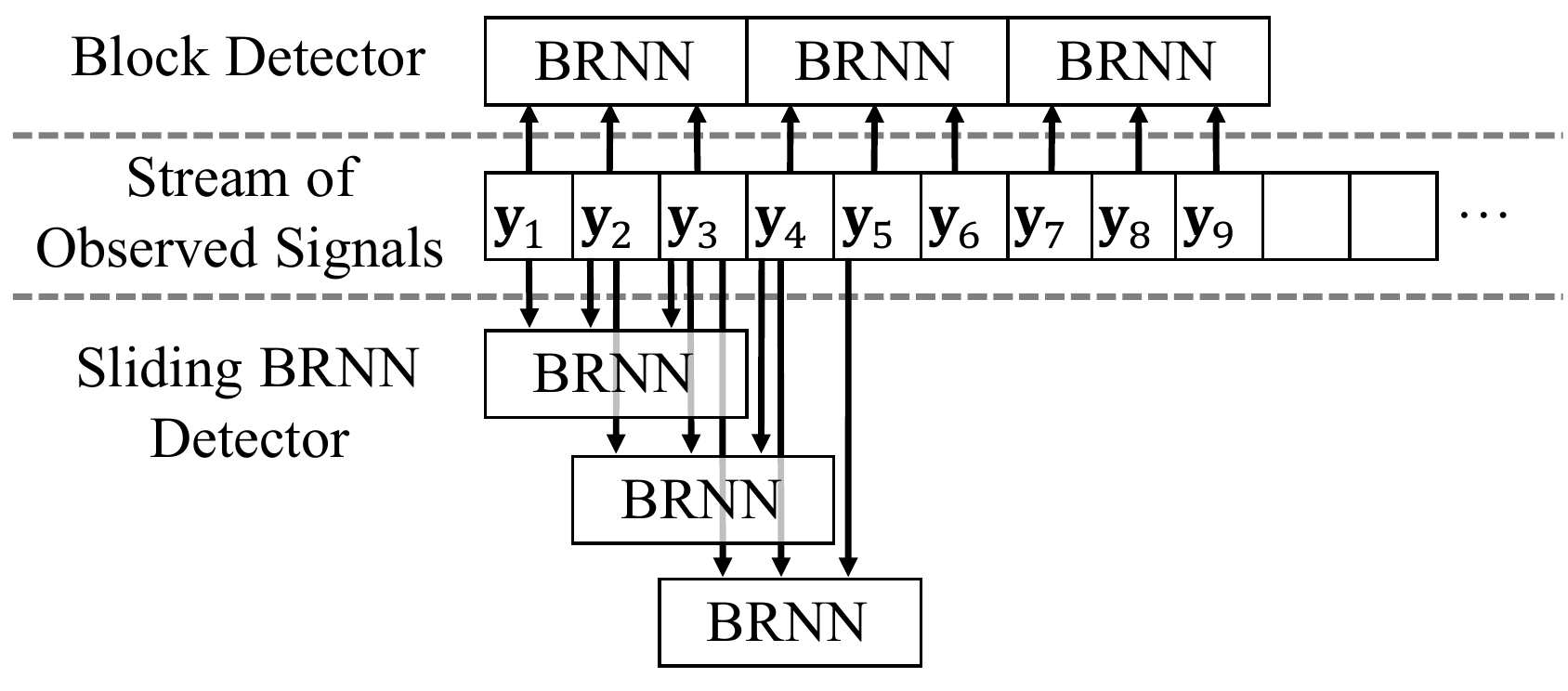}
	\caption{\label{fig:slidingDetector} The sliding BRNN detector.}
	\vspace{-0.3cm}
\end{figure}

First, we fix the length of the BRNN. Ideally, the length must be the same size as the memory length of the channel. However, if this is not known in advance, the BRNN length can be treated as a hyper parameter to be tuned during training. Let $L$ be the length of the BRNN. Then during training, all blocks of $L$ consecutive transmissions are used for training. After training, the simplest scheme would be to detect the stream of incoming data in blocks of length $L$ as shown in the top portion of Figure~\ref{fig:slidingDetector}. The main drawback here is that the symbols at the end of each block may affect the symbols in the next block and this relation is not captured in this scheme. Another issue is that $L$ consecutive symbols must be received before detection can be preformed.

To overcome these limitations, inspired by some of the techniques used in speech recognition \cite{gra06}, we propose a scheme we call the {\em sliding BRNN (SBRNN) detector}. The first $L$ symbols are detected using the BRNN. Then as each new symbol arrives at the destination, the position of the BRNN is slided ahead by one symbol. Let the set $\mathcal{J}_k = \{j \mid j \leq k ~\wedge~ j+L > k \}$ be the set of all valid staring positions for a BRNN detector of length $L$, such that the detector overlaps with the \kth~symbol. For example, if $L=3$ and $k=4$, then $j=1$ is not in the set $\mathcal{J}_k$ since the BRNN detector overlaps with symbol positions $1,2,3$ and not the symbol position 4. Let $\hat{\vec{p}}^{(j)}_k$ be the estimated PMF for the \kth~symbol, when the start of the sliding BRNN is on $j\in\mathcal{J}_k$. The final PMF corresponding to the \kth~symbol is given by averaging the estimated PMFs for each of the relevant windows:
\begin{align}
	\label{eq:slidingPMF}
	\hat{\vec{p}}_k = \frac{1}{|\mathcal{J}_k|} \sum_{j\in \mathcal{J}_k} \hat{\vec{p}}^{(j)}_k.
\end{align}
One of the main benefits of this approach is that after the first $L$ symbols are received and detected, as the signal corresponding to a new symbol arrives at the destination, the detector immediately estimates that symbol. The detector also updates its estimate for the previous $L-1$ symbols dynamically. Therefore, this algorithm is similar to a dynamic programming algorithm.

%
%

The bottom portion of Figure~\ref{fig:slidingDetector} illustrates the sliding BRNN detector. In this example, after the first 3 symbols arrive, the PMF for the first three symbols, $i\in\{ 1,2,3\}$, is given by $\hat{\vec{p}}_i = \hat{\vec{p}}^{(1)}_i$. When the 4th symbol arrives, the estimate of the first symbol is unchanged, but for $i\in\{2,3\}$, the second and third symbol estimates are updated as $\hat{\vec{p}}_i = \tfrac{1}{2}(\hat{\vec{p}}^{(1)}_i+\hat{\vec{p}}^{(2)}_i)$, and the 4th symbol is estimated by $\hat{\vec{p}}_4 = \hat{\vec{p}}^{(2)}_4$. Note that although in this paper we assume that the weights of all $\hat{\vec{p}}^{(j)}_k$ are $\tfrac{1}{|\mathcal{J}_k|}$, the algorithm can use different weights. Moreover, the complexity of SBRNN increases linearly with the length of the BRNN.

To evaluate the performance of all these deep learning detection algorithms, we use an experimental platform to collect data for training and testing the algorithms. 

\section{Experimental Setup}
\label{sec:expSetup}

The in-vessel molecular communication experimental platform, presented in \cite{far17Expt}, is used for evaluation. The platform uses peristaltic pumps to inject different chemicals into a main fluid flow in small silicon tubes. Multiple tubes with different diameters can be networked in branches to replicate a more complex environment such as the cardiovascular system in the body or complex networks of pipes in industrial complexes and city infrastructures. 
In our platform, the main fluid flow is water and the transmitter used acids (vinegar) and bases (window cleaning solution) to encode information on the pH level. We use these particular chemicals since they are easily available and inexpensive. However, the results can be extended to blood as the main flow, and proteins and enzymes as the chemicals that are released by the transmitter.

\subsection{System Model}
The field of molecular communication is relatively new, and thus, the models that have been developed are not validated experimentally yet. In fact, there are no standardized models for these systems and most previous work have used the diffusion equation to model the chemical propagation from the sender to the destination \cite{eckBook,far16ST}. However, these works consider only a single chemical species in the environment.

In many molecular systems multiple chemical species are present, and therefore, chemical interactions will be an integral part of the system. In fact in a closed system, using multiple types of reactive chemicals can be beneficial since transmitting a single chemical repeatedly will saturate/contaminate the environment and degrades the system performance. That is the motivation behind using multiple reactive chemicals such as acids and bases for transmission. When chemical reactions are present in the system, the average behavior of signal propagation can be represented by a system of PDEs known as reaction-diffusion equations, which are difficult to solve analytically. For our acid-base platform, the system of PDEs are given by
	\begin{align}
		\label{eq:reactDiffEqH}
		\frac{\partial C_{H}}{\partial t} &=   D_{H} \nabla^2 C_{H}-\nabla.(\mathbf{v}C_{H})-k_f C_{H} C_{OH}+k_r  \\
		\label{eq:reactDiffEqOH}
		\frac{\partial C_{OH}}{\partial t} &=   D_{OH} \nabla^2 C_{OH}-\nabla.(\mathbf{v}C_{OH})-k_f C_{H} C_{OH}+k_r,
	\end{align}
where $C_{H}$ and $C_{OH}$ are the concentration of acid and base, respectively, $D_{H}$ and $D_{OH}$ are the diffusion coefficient of the acid and base, and $k_f$ and $k_r$ are the reaction rates of the system. 
This system of PDE is extremely difficult to solve, and we have not been able to find any approximate solution for it. Note that even if a solution was obtained, it only models the average behavior of the signal propagation and does not model the stochastic nature of the transport. Moreover, models for the perturbations introduced by the transmission and reception processes are also required. Since all of these problems makes a traditional model based approach to system design challenging, a new data driven approach using deep learning would be a suitable alternative.

\subsection{Detection}  
In the platform, time-slotted communication is employed where the transmitter modulates information on acid and base signals by injecting these chemicals into the channel during each symbol duration.
We use a binary modulation in this work where the 0-bit is transmitted by pumping acid into the environment for 30 ms at the beginning of the symbol interval, and the 1-bit is represented by pumping base into the environment for 30 ms at the beginning of the symbol interval. The symbol interval consists of this 30 ms injection interval followed by a period of silence, which can also be considered as a guard band between symbols. In particular, four different silence durations (guard bands) of 220 ms, 304 ms, 350 ms, and 470 ms are used in this work to represent bit rates of 4, 3, 2.6, and 2 bps.

\begin{figure}
	\centering
	\includegraphics[width=0.7\columnwidth,keepaspectratio]{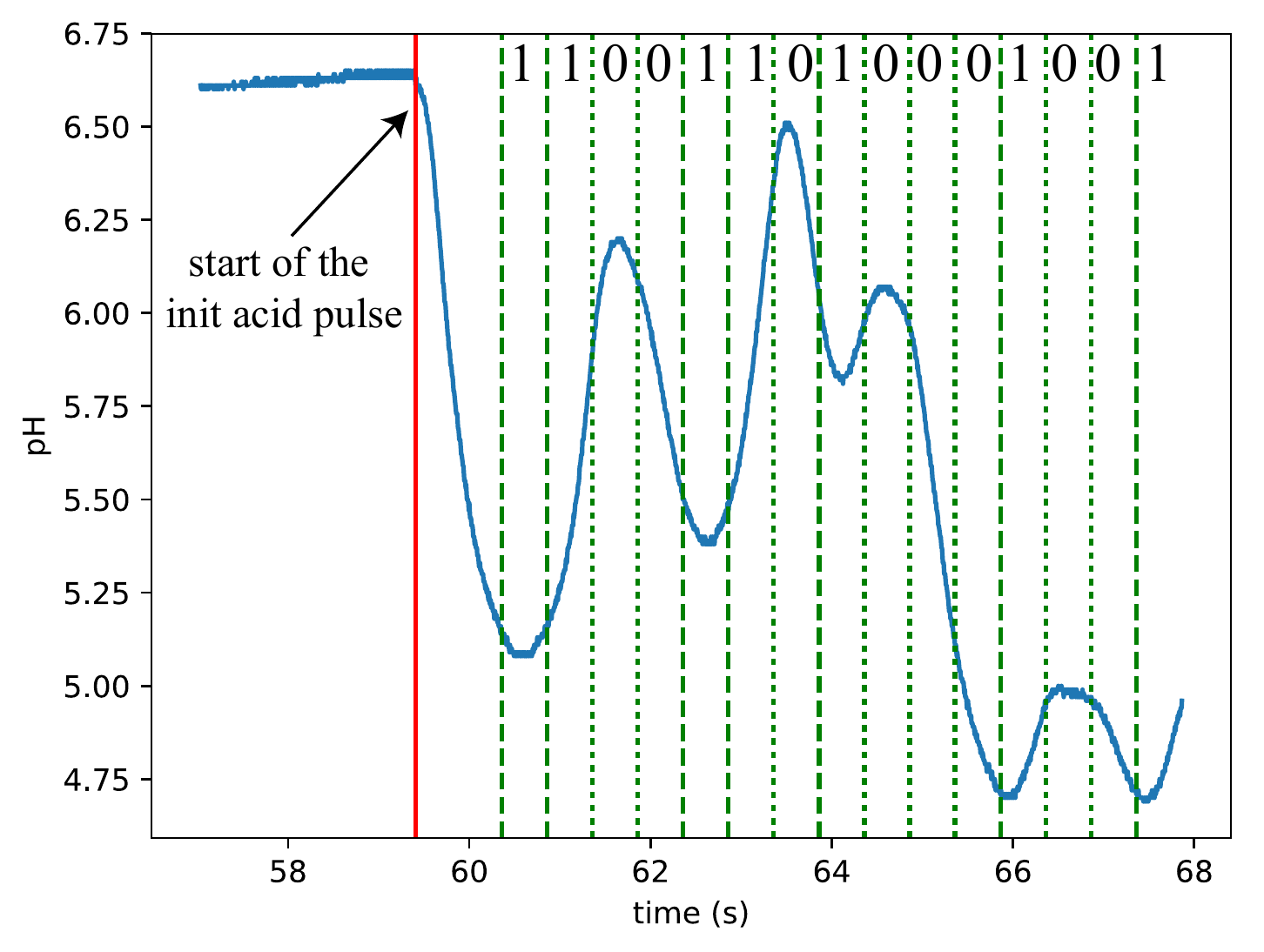}
	\vspace{-0.3cm}
	\caption{\label{fig:sampleTrans} Sample received pH signal.}
	\vspace{-0.3cm}
\end{figure}
To synchronize the transmitter and the receiver, every message sequence starts with one initial injection of acid into the environment for 100 ms followed by 900 ms of silence. The receiver then detects the starting point of this pulse and uses it to synchronize with the transmitter. Figure~\ref{fig:sampleTrans} shows the received pH signal for the transmission sequence ``110011010001001''. The start of the initial acid pulse detected by the receiver is shown using the red line. This detected time is used for synchronization and all the subsequent symbol intervals are shown by the green dashed and dotted lines. The dashed lines are used to indicate a 1-bit transmission and dotted lines to indicate a 0-bit. 

\begin{figure}
	\vspace{-0.1cm}
	\centering
	\includegraphics[width=0.7\columnwidth,keepaspectratio]{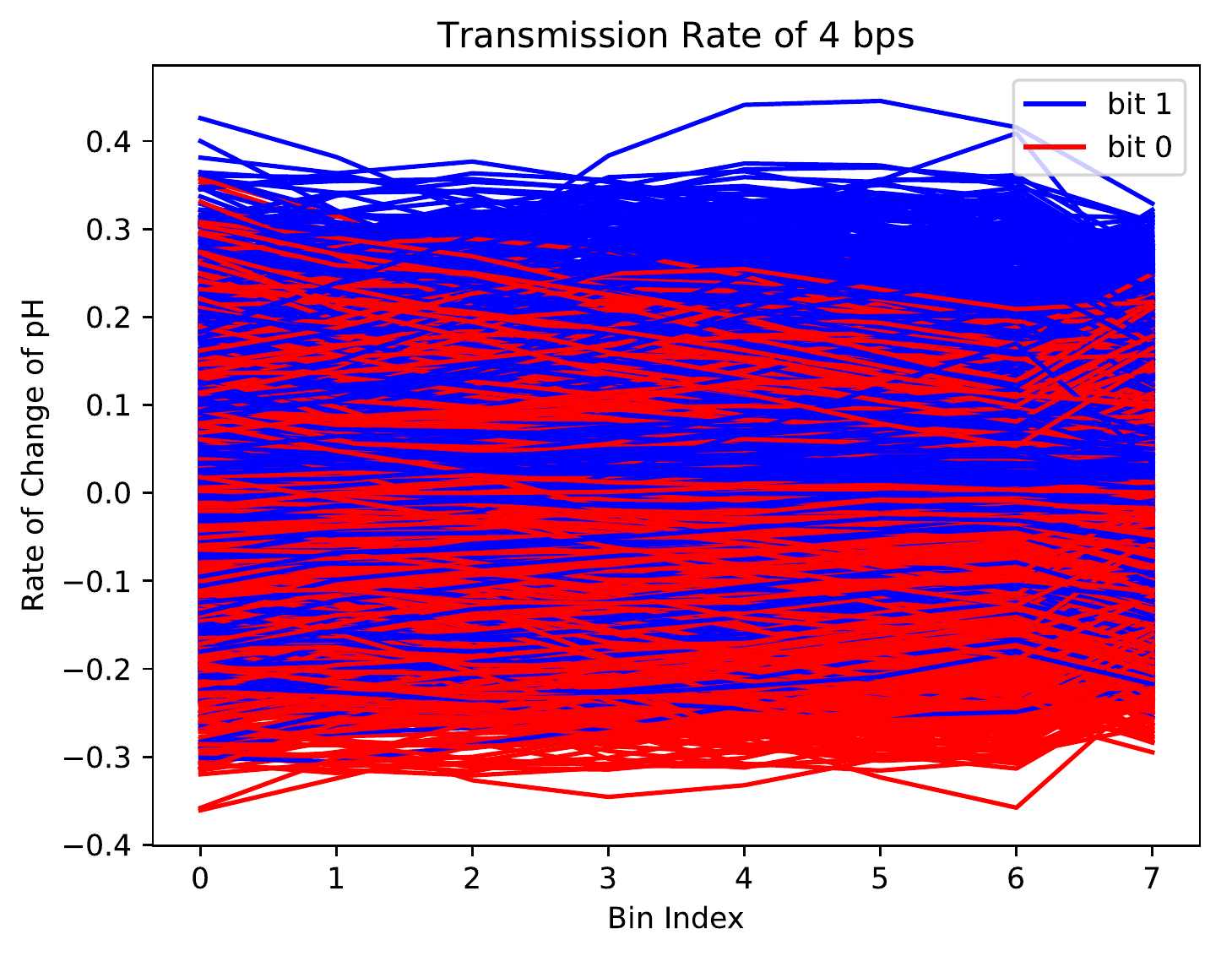}
	\vspace{-0.3cm}
	\caption{\label{fig:RCeye4bps} Experimental data for rate of change $\vec{d}$ for $B=9$.}
	\vspace{-0.3cm}
\end{figure}

Although it is difficult to obtain analytical models for multi-chemical communication systems as explained in the previous section, it is expected that when an acid pulse is transmitted, the pH should drop, and when a base pulse is injected into the environment, the pH should increase. Therefore, one approach to detection is to use the rate of change of pH to detect the symbols. Note that rate of change of concentration is used in previous experimental demonstrations of chemical communication \cite{far13,koo16}, and because there were no models that matched our experimental data, it is the only technique which we can compare our deep learning detectors against. 

To remove the noise from the raw pH signal, we divide the symbol interval (the time between green lines in  Figure~\ref{fig:sampleTrans}) into a number of equal subintervals or bins. Then the pH values inside each bin are averaged to represent the pH value for the corresponding bin. Let $B$ be the number of bins, and $\vec{b}=[b_0,b_1,\cdots,b_{B-1}]$ the corresponding values of each bin. The difference between values of these bins $\vec{d}=[d_0,d_1,\cdots,d_{B-2}]$, where $d_{i-1} = b_i- b_{i-1}$, are used as a baseline detection algorithm. This algorithm has two parameters: the number of bins $B$, and the index $\gamma$ that is used for detection. If $d_\gamma\leq0$, acid transmission and hence the 0-bit is detected, and if $d_\gamma>0$, the 1-bit is detected. Figure~\ref{fig:RCeye4bps} shows the $\vec{d}$ values of some sample experimental data for $B=9$. In the plot, the 1-bit symbols are shown in blue and the 0-bit in red. From the figure it is evident that detecting the bits is challenging due to chemical interactions. 



\spaceup
\section{Results}
\label{sec:results}
\spaceup
We use our experimental platform to collect measurement data and create the dataset that is used for training and testing the detection algorithms. Particularly, as explained in the previous section, four symbol durations of 250 ms, 334 ms, 380 ms and 500 ms are considered which results in data rates ranging from 2 to 4 bits per second (bps). For each symbol interval, random bit sequences of length 120 are transmitted 100 times, where each of the 100 transmissions are separated in times. Since we assume no channel coding is used, the bits are iid and equiprobable.  This results in 12k bits per symbol duration that is used for training and testing. From the data, 84 transmissions per symbol duration (10,080 bits) are used for training and 16 transmissions are used for testing (1,920 bits). Therefore, the total number of training bits are 40,320, and the total number of bits used for testing is 7,680. Although the dataset is not large because collecting experimental measurements is laborious, training with larger datasets is demonstrated in the extension of this work~\cite{far18TSP}.

We start by considering the baseline detection using the rate of change of the pH. We use the training data to find the best detection parameters $B$ and $\gamma$, and the test data for evaluating the performance. Besides this algorithm we consider different deep learning detectors. For all training, the Adam optimization algorithm \cite{kin14} is used with the learning rate $10^{-3}$.  Unless specified otherwise the number of epoch used during training is 200 and the batch size is 10. All the hyper parameters are tuned using grid search.

We use two symbol-by-symbol detectors based on deep learning. The first detector uses three fully connected layers with 80 hidden nodes and a final softmax layer for detection. Each fully connected layer uses the rectified linear unit (ReLU) activation function. The input to the network are a set of features extracted from the received signal, which are chosen based on performance and the characteristics of the physical channel. The input includes: $b_1$ and $b_B$, i.e., the pH level in the first and the last bins, $\vec{d}$, i.e., the vector of differences of consecutive bins, and a number that indicates the symbol duration. Here, we refer to this network as {\em Base-Net}. A second symbol-by-symbol detector uses 1-dimensional CNNs. Particularly, the best network architecture that we found has the following layers. 1) 16 filters of length 2 with ReLU activation; 2) 16 filters of length 4 with ReLU activation; 3) max pooling layer with pool size 2; 4) 16 filters of length 6 with ReLU activation; 5) 16 filters of length 8 with ReLU activation; 6) max pooling layer with pool size 2; 7) flatten and a softmax layer. The stride size for the filters is 1 in all layers. The input to this network is the vector of pH values corresponding to each bin $\vec{b}$. We refer to this network as {\em CNN-Net}.

For the sequence detection, we use three networks, two based on RNNs and one based on the SBRNN. The first network has 3 LSTM layers and a final softmax layer, where the length of the output of each LSTM layer is 40. Two different inputs are used with this network. In the first, the input is the same set of features as the Base-Net above. We refer to this network as {\em LSTM3-Net}. In the second, the input is the pretrained CNN-Net described above without the top softmax layer. In this network, the CNN-Net chooses the features directly from the pH levels of the bins. We refer to this network as {\em CNN-LSTM3-Net}. Finally, we consider three layers of bidirectional LSTM cells, where each cell's output length is 40, and a final softmax layer. The input to this network are the same set of features used for Base-Net and the LSTM3-Net. When this network is used, during testing we use the SBRNN algorithm. We refer to this network as {\em SBLSTM3-Net}. For all the sequence detection algorithms, during testing, sample data sequences of the 120 bits are treated as an incoming data stream, and the detector estimates the bits one-by-one, simulating a real communication scenario.

We have trained each network using different number of bins $B$ to find the best value for each network. For the Base-Net $B=9$, for the CNN-Net $B=30$ and for all networks where the first layer is an LSTM or a BLSTM cell $B=8$. Note that during the training, for all deep learning detectors, the data from all symbol durations are used to train a single network, which can then perform detection on all symbol durations.
\begin{table}[t]
	\scriptsize 
	\caption{Bit Error Rate Performance}
	\vspace{-0.3cm}
	\label{tb:BER}
	\centering
	\begin{tabular}{c|cccc}
		\toprule
		Symb. Dur. 			& 250 ms 		& 334 ms 		& 380 ms 		& 500 ms  \\
		\midrule
		Baseline			& 0.1297		& 0.0755  		& 0.0797 		& 0.0516  \\
		Base-Net   			& 0.1057		& 0.0245  		& 0.0380 		& 0.0115  \\
		CNN-Net    			& 0.1068    	& 0.0750  		& 0.0589 		& 0.0063  \\
		CNN-LSTM3-Net120	& 0.0677		& 0.0271		& 0.0026 		& 0.0021  \\
		LSTM3-Net120		&{\bf 0.0333}	& 0.0417		& 0.0083 		& 0.0005 \\
		SBLSTM3-Net10		& 0.0406		& {\bf 0.0141}	& {\bf 0.0005}	& {\bf 0.0000} \\
		\bottomrule
	\end{tabular}
	\vspace{-0.6cm}
\end{table}
%

Table \ref{tb:BER} summarizes the best BER performance we obtain for all detection algorithms, including the baseline algorithm, by tuning all the hyper parameters using grid search. The number in front of the sequence detectors, indicates the sequence length. For example, LSTM3-Net120 is an LSTM3-Net that is trained on 120 bit sequences. In general, algorithms that use sequence detection perform significantly better than any symbol-by-symbol detection algorithm including the baseline algorithm. This is due to significant ISI present in chemical communication systems. Overall, the proposed SBLSTM algorithm performs better than all other NN detectors considered. Note that BER values below $5\times 10^{-3}$ are not very accurate since the number of errors in the test dataset are less than 10, and more errors would be required for a better estimation of BER.

To demonstrate the practicality of the proposed scheme, we implement the trained deep learning detectors as part of a text messaging service on the platform\footnote{A video of this text messaging service and the deep learning detector detecting the bits in real-time can be viewed at \url{http://narimanfarsad.com/pH-setup.mp4}.}. The text message could be of any length, and we are able to reliably transmit and receive messages at 2 bps. This data rate is an order of magnitude higher than previous systems \cite{far13,koo16}.

\spaceup
\section{Conclusions}
\spaceup
We used several deep learning architectures for building detectors for communication systems. Different architectures were considered for symbol-by-symbol detection as well as sequence detection. We also proposed a new sequence detection scheme called sliding bidirectional recurrent network (SBRNN). These algorithms could be used in systems where the underlying physical models of the channel are unknown or inaccurate. We use an experimental platform that simulates in-vessel chemical communication to collect experimental data for training and testing deep learning algorithms. We show that deep learning sequence detectors can improve the detection performance significantly compared to a baseline approach used in previous works \cite{far13,koo16}. Moreover, in a journal extension of this work, we use a Poisson channel model for molecular communication systems with none-reactive chemicals, to show that the performance of the proposed SBRNN detector can be close to an optimal Viterbi detector in low noise environments. We also demonstrate that SBRNN is resilient to changing channel conditions and can perform detection without channel state information. These demonstrate the promising performance deep learning detection algorithms could have in designing some of the future communication systems. 

\bibliographystyle{IEEEbib}
\bibliography{MolCom-Nariman,ML-Nariman}

\begin{thebibliography}{10}

\bibitem{sto09}
M.~Stojanovic and J.~Preisig,
\newblock ``Underwater acoustic communication channels: Propagation models and
  statistical characterization,''
\newblock {\em IEEE Communications Magazine}, vol. 47, no. 1, pp. 84--89, 2009.

\bibitem{mor06}
Y.~Moritani, S.~Hiyama, and T.~Suda,
\newblock ``Molecular communication for health care applications,''
\newblock in {\em Proc. of 4th Annual IEEE International Conference on
  Pervasive Computing and Communications Workshops}, Pisa, Italy, 2006, p.~5.

\bibitem{aky08}
Ian~F. Akyildiz, Fernando Brunetti, and Cristina Blazquez,
\newblock ``Nanonetworks: A new communication paradigm,''
\newblock {\em Computer Networks}, vol. 52, no. 12, pp. 2260--2279, August
  2008.

\bibitem{eckBook}
Tadashi Nakano, Andrew~W. Eckford, and Tokuko Haraguchi,
\newblock {\em Molecular communication},
\newblock Cambridge University Press, 2013.

\bibitem{far16ST}
N.~Farsad, H.~B. Yilmaz, A.~Eckford, C.~B. Chae, and W.~Guo,
\newblock ``A comprehensive survey of recent advancements in molecular
  communication,''
\newblock {\em IEEE Communications Surveys \& Tutorials}, vol. 18, no. 3, pp.
  1887--1919, thirdquarter 2016.

\bibitem{lec15}
Yann LeCun, Yoshua Bengio, and Geoffrey Hinton,
\newblock ``Deep learning,''
\newblock {\em Nature}, vol. 521, no. 7553, pp. 436--444, May 2015.

\bibitem{goodfellowBook}
Ian Goodfellow, Yoshua Bengio, and Aaron Courville,
\newblock {\em Deep {Learning}},
\newblock MIT Press, Nov. 2016.

\bibitem{ibn00}
Mohamed Ibnkahla,
\newblock ``Applications of neural networks to digital communications – a
  survey,''
\newblock {\em Signal Processing}, vol. 80, no. 7, pp. 1185--1215, 2000.

\bibitem{aaz92}
B.~Aazhang, B.~P. Paris, and G.~C. Orsak,
\newblock ``Neural networks for multiuser detection in code-division
  multiple-access communications,''
\newblock {\em IEEE Transactions on Communications}, vol. 40, no. 7, pp.
  1212--1222, Jul 1992.

\bibitem{mit94}
U.~Mitra and H.~V. Poor,
\newblock ``Neural network techniques for adaptive multiuser demodulation,''
\newblock {\em IEEE Journal on Selected Areas in Communications}, vol. 12, no.
  9, pp. 1460--1470, Dec 1994.

\bibitem{jua06}
Juan~J. Murillo-fuentes, Sebastian Caro, and Fernando P\'{e}rez-Cruz,
\newblock ``Gaussian processes for multiuser detection in cdma receivers,''
\newblock in {\em Advances in Neural Information Processing Systems 18},
  Y.~Weiss, P.~B. Sch\"{o}lkopf, and J.~C. Platt, Eds., pp. 939--946. MIT
  Press, 2006.

\bibitem{isi07}
Yal{\c{c}}{\i}n I{\c{s}}{\i}k and Necmi Ta{\c{s}}p{\i}nar,
\newblock ``Multiuser detection with neural network and pic in cdma systems for
  awgn and rayleigh fading asynchronous channels,''
\newblock {\em Wireless Personal Communications}, vol. 43, no. 4, pp.
  1185--1194, 2007.

\bibitem{nac16}
E.~Nachmani, Y.~Be'ery, and D.~Burshtein,
\newblock ``Learning to decode linear codes using deep learning,''
\newblock in {\em 2016 54th Annual Allerton Conference on Communication,
  Control, and Computing (Allerton)}, Sept 2016, pp. 341--346.

\bibitem{osh16}
T.~J. O'Shea, K.~Karra, and T.~C. Clancy,
\newblock ``Learning to communicate: Channel auto-encoders, domain specific
  regularizers, and attention,''
\newblock in {\em 2016 IEEE International Symposium on Signal Processing and
  Information Technology (ISSPIT)}, Dec 2016, pp. 223--228.

\bibitem{lee17}
Changmin Lee, H.~Birkan Yilmaz, Chan-Byoung Chae, Nariman Farsad, and Andrea
  Goldsmith,
\newblock ``Machine learning based channel modeling for molecular {MIMO}
  communications,''
\newblock in {\em IEEE International Workshop on Signal Processing Advances in
  Wireless Communications (SPAWC)}, 2017.

\bibitem{far17Expt}
Nariman Farsad, David Pan, and Andrea Goldsmith,
\newblock ``A novel experimental platform for in-vessel multi-chemical
  molecular communications,''
\newblock in {\em IEEE Global Communications Conference (GLOBECOM),}, 2017,
\newblock accepted.

\bibitem{far13}
N.~Farsad, W.~Guo, and A.~W. Eckford,
\newblock ``Tabletop molecular communication: Text messages through chemical
  signals,''
\newblock {\em PLOS ONE}, vol. 8, no. 12, pp. e82935, Dec 2013.

\bibitem{koo16}
B.~H. Koo, C.~Lee, H.~B. Yilmaz, N.~Farsad, A.~Eckford, and C.~B. Chae,
\newblock ``Molecular {MIMO}: From theory to prototype,''
\newblock {\em IEEE Journal on Selected Areas in Communications}, vol. 34, no.
  3, pp. 600--614, March 2016.

\bibitem{hin12}
G.~Hinton, L.~Deng, D.~Yu, G.~E. Dahl, A.~r.~Mohamed, N.~Jaitly, A.~Senior,
  V.~Vanhoucke, P.~Nguyen, T.~N. Sainath, and B.~Kingsbury,
\newblock ``Deep neural networks for acoustic modeling in speech recognition:
  The shared views of four research groups,''
\newblock {\em IEEE Signal Processing Magazine}, vol. 29, no. 6, pp. 82--97,
  2012.

\bibitem{bah14}
Dzmitry Bahdanau, Kyunghyun Cho, and Yoshua Bengio,
\newblock ``Neural {Machine} {Translation} by {Jointly} {Learning} to {Align}
  and {Translate},''
\newblock {\em arXiv:1409.0473 [cs, stat]}, Sept. 2014.

\bibitem{li16}
Zhen Li and Yizhou Yu,
\newblock ``Protein {Secondary} {Structure} {Prediction} {Using} {Cascaded}
  {Convolutional} and {Recurrent} {Neural} {Networks},''
\newblock {\em arXiv:1604.07176 [cs, q-bio]}, 2016.

\bibitem{hoc97}
Sepp Hochreiter and J{\"u}rgen Schmidhuber,
\newblock ``Long short-term memory,''
\newblock {\em Neural computation}, vol. 9, no. 8, pp. 1735--1780, 1997.

\bibitem{sch97}
Mike Schuster and Kuldip~K Paliwal,
\newblock ``Bidirectional recurrent neural networks,''
\newblock {\em IEEE Transactions on Signal Processing}, vol. 45, no. 11, pp.
  2673--2681, 1997.

\bibitem{gra05}
Alex Graves and J{\"u}rgen Schmidhuber,
\newblock ``Framewise phoneme classification with bidirectional lstm and other
  neural network architectures,''
\newblock {\em Neural Networks}, vol. 18, no. 5, pp. 602--610, 2005.

\bibitem{gra06}
Alex Graves, Santiago Fernández, and Faustino Gomez,
\newblock ``Connectionist temporal classification: Labelling unsegmented
  sequence data with recurrent neural networks,''
\newblock in {\em In Proceedings of the International Conference on Machine
  Learning, ICML 2006}, 2006, pp. 369--376.

\bibitem{far18TSP}
Nariman Farsad and Andrea Goldsmith,
\newblock ``Neural network detectors for sequence detection in communication
  systems,''
\newblock {\em IEEE Transactions on Signal Processing}, 2018,
\newblock submitted, [Online]. Available: https://arxiv.org/abs/1802.02046.

\bibitem{kin14}
Diederik Kingma and Jimmy Ba,
\newblock ``Adam: A method for stochastic optimization,''
\newblock {\em arXiv preprint arXiv:1412.6980}, 2014.

\end{thebibliography}

\end{document}